\documentclass[usenatbib]{mn2e}
\usepackage{graphicx,lscape}

\begin{document}

\title
[Age Effect on Ca Triplet]
{The Effects of Age on Red Giant
Metallicities Derived
From the Near-Infrared
Ca II Triplet\thanks{Based on data obtained
at the Very Large Telescope of the European
Southern Observatory's Paranal Observatory,
under programme 70.B-0398.}}

\author
[A.A. Cole et al.]
{A.A. Cole,$^1$\thanks{{\it cole@astro.rug.nl}}
 T.A. Smecker-Hane,$^2$ E. Tolstoy,$^1$
 T.L. Bosler$^2$ and J.S. Gallagher,$^3$ \\
$^1$Kapteyn Astronomical Institute, University of Groningen,
 Postbus 800, 9700 AV Groningen, The Netherlands. \\
$^2$Department of Physics and Astronomy, University of California,
 Irvine, 4129 Frederick Reines Hall, Irvine, CA 92697-4575. \\
$^3$Department of Astronomy, University of Wisconsin-Madison,
 5534 Sterling Hall, 475 North Charter Street, Madison, WI 53706-1582. \\
}

\maketitle
\begin{abstract}
We have obtained spectra with resolution $\sim$2.5 \AA\ in 
the region $\approx$ 7500--9500 \AA\ for 116 red giants
in 5 Galactic globular clusters and 6 old open clusters
(5 with published metallicities, and 1 previously unmeasured).
The signal-to-noise ranges from 20 $\leq$ S/N $\leq$ 85. 
We measure the equivalent widths of the infrared Ca~II
triplet absorption lines in each star and compare to 
cluster metallicities taken from the literature.
With globular cluster abundances
on the Carretta \& Gratton scale, and open cluster abundances
taken from the compilation of Friel and collaborators, we
find a linear relation between [Fe/H] and Ca~II line strength
spanning the range $-$2 $\la$ [Fe/H] $\la$ $-$0.2
and 2.5 $\la$ (age/Gyr) $\la$ 13.  The reference
abundance scales appear to be consistent with each other
at the $\sim$0.1 dex level.  Alternate choices for
metallicity scales can introduce curvature into
the relation between [Fe/H] and Ca~II equivalent width.
No evidence for an age effect on the metallicity
calibration is observed.
Using this calibration, we find the metallicity of the
massive, old, open cluster Trumpler~5 to be 
[Fe/H] = $-$0.56 $\pm$0.11.  This is the first
spectroscopic abundance measurement for Trumpler~5, and is
lower by $\approx$ 0.3 dex than estimates based on the
cluster colour-magnitude diagram.  Considering the 10
clusters of known metallicity shifted to a common distance
and reddening, we find that the 
additional error introduced by the variation of
horizontal branch/red clump magnitude with metallicity
and age is of order $\pm$0.05 dex, which can be neglected
in comparison to the intrinsic scatter in the method.
The results are discussed in the context of abundance
determinations for red giants in Local Group galaxies.
\end{abstract}


\section{Introduction}

One of the most widely applied techniques for the 
derivation of abundances of individual red giants
in very distant systems (globular clusters or 
external galaxies) is the measurement of the
strength of the near-infrared lines of
Ca~{\small II} at $\lambda \lambda$ = 
8498.02, 8542.09, 8662.14 \AA.  
As noted in pioneering papers by \citet{arm88}
and \citet{arm91},
the calcium triplet (CaT) is attractive for several reasons:
the red giants are near their brightest in 
this wavelength range; the lines are so broad
that they can be measured at moderate resolution;
they arise between excited states of the Ca$^+$ ion
and so are relatively unaffected by interstellar
absorption;
and they lie in a window of small telluric 
absorption between H$_2$O and O$_2$ bands.
Ca~{\small II} is the dominant ionization stage
in red giant branch (RGB) stars, and because the
continuous opacity in the photospheres is
dominated by the H$^-$ ion, the relation 
between line strength and surface gravity
is theoretically well-modelled \citep{jor92}.
Additionally, the CaT is attractive because calcium abundances
are not expected to be affected by proton-capture nucleosynthesis
in the intermediate- and low-mass stars of the RGB.  This
distinguishes it from lighter elements such as O, Na, Mg, Al,
and Si \citep[e.g.,][]{iva01}.  Thus the amount of calcium
in an RGB envelope should be representative of the primordial
abundances in the star.  

However, the matter is complicated by the very
strength of the lines that makes them attractive
in the first place.  The CaT lines are strongly
saturated and extremely broad.  The 
line strength thus depends strongly on the 
temperature structure of the upper photosphere
and chromosphere, regions which are notoriously
difficult to model in red giants.  Ca is 
an $\alpha$-element, so its abundance relative
to the iron-peak elements is expected to vary
with environment based on the relative
importance of Type~I and Type~II supernovae to
chemical evolution.  Quantifying the impact of [$\alpha$/Fe]
variation on abundances derived from the CaT
is nontrivial, because the $\alpha$-enhancement
affects both the stellar temperature and the
electron density in the atmosphere.  A detailed
theoretical understanding of the relation
between CaT line strength, metallicity, and
the evolutionary state of a giant star is 
quite challenging.  

In contrast, a large body of empirical work
shows that the CaT line strength is highly
sensitive to overall stellar metallicity
\citep[e.g.,][hereafter R97b]{rut97b}.
Their calibration uses a simple linear
relation between the CaT pseudo-equivalent
widths and V magnitudes of individual red 
giants to effectively remove the T$_{eff}$
and log~$g$ dependencies.  A spectrophotometric
index that is sensitive only to metallicity
is thus defined and calibrated to a sample
of known metallicity.
Qualitative theoretical understanding
of this line strength behavior is based on the
response to changes induced in the stellar
atmosphere structure by metallicity variations
\citep{jor92}.  

Because the CaT index is empirically calibrated,
it is potentially dangerous to apply the technique
to stellar populations beyond the range of the
calibration sample.  The most complete such 
calibration to date has been published by
\citet[][hereafter R97a]{rut97a} and R97b,
who found a linear relation between the 
reduced CaT equivalent widths and 
metallicities on the high-dispersion globular
cluster abundance scale of 
\citet[][hereafter CG97]{car97}.
This calibration, while extremely useful, only covers
the metallicity range from $-$2.1 $\la$ [Fe/H]
$\la$ $-$0.6 and the age range from 9 $\la$
(age/Gyr) $\la$ 13.  
This is currently one of the limiting
uncertainties in understanding the age-metallicity
relation and abundance gradient of the Large Magellanic
Cloud (LMC) \citep{col00,sme03}, for example,
and might be problematic for planned
studies of mixed-age and metal-rich systems like the M31 halo.

It is important to develop a detailed understanding
of the calcium triplet technique because at present
it is the {\it only} way to efficiently 
make spectroscopic abundance assessments 
of red giants in galaxies beyond the 
Milky Way halo; this makes it an essential
observational tool for the understanding of
chemical evolution of galaxies.
Here, we show that the age effect 
is a negligible source of error for metallicities
derived from the CaT index.  One of us (TLB)
will expand on this result in a forthcoming analysis
of high- and low-dispersion spectra of numerous 
Galactic star
cluster giants, producing a better understanding
of the relationship between the calcium abundance
and the CaT line strength as part of her Ph.~D.~thesis.

In this paper we move towards the calibration of
the CaT index as a metallicity indicator for individual
red giants at younger ages than 
the globular clusters that define the present CaT
abundance scale.   
The following section describes the
sample of observed clusters, the data reduction process,
and the measurement of radial velocities and equivalent
widths.  Section \ref{cal-sec} discusses the relation between calcium
line strength and metallicity for the 10 clusters of known
abundance, and the effects of age on the metallicity 
measurement.  In \S \ref{tr5-sec}, we apply our calibration of the
CaT index to the little-studied massive cluster Trumpler 5,
producing the first spectroscopic metallicity assessment for
the cluster.  We consider the systematic errors introduced
by taking a stellar sample of mixed age and metallicity,
as in Local Group galaxies in section \ref{mix-sec}.
Lastly, we briefly discuss the choice of reference metallicity
scales and their effects on our calibration.

\section{Data: Observations \& Reductions}
\label{obs-sec}

The data were obtained under ESO programme 70.B-0398,
a study of the metallicity distribution function of the
central bar of the LMC.  
Because most LMC field stars are much younger
than the Galactic globular clusters
\citep{but77,sme02} and have metallicities near the upper end
of the R97b calibration
\citep{col00,sme03}, we observed a sample of five globular
clusters and six old open clusters to calibrate our LMC
sample and to begin the extension of the CaT abundance 
scale to younger ages and higher metallicities.
The observed clusters and their relevant physical
parameters are summarized in Table \ref{clus-tab}.

\subsection{Target Selection}

We chose to observe globular clusters that have
already been extensively used in the calibration of the 
CaT \citep[e.g.][]{sme03,tol01,col00,rut97a,gei95,dac95,
sun93,arm91,ols91},
limited to those with good late-December visibility
from northern Chile.  All of these clusters, except NGC 1851,
have had high-dispersion metallicity analyses published
by CG97;  NGC 1851 has been measuered with the CaT technique
and calibrated to the CG97 scale
by R97b.  While the absolute ages of globular
clusters remain uncertain, our sample has been
placed onto a consistent scale by \citet{sal02}.  
Table \ref{glob-tab} shows the individual red giants 
observed in each cluster, with the adopted V magnitude,
derived radial velocities and CaT equivalent widths,
and references to the identifications and photometry.
The number of stars observed in each cluster depends on the
cluster angular size and density and the spectrograph field
geometry, hence relatively few stars have been observed
in nearby (47 Tuc) or sparse (NGC 2298) clusters.

Open cluster targets were chosen from the 
\citet[][hereafter F02]{fri02}
list of old clusters with metallicity estimates based on
medium-resolution measurements of iron and iron-peak-element
line strengths.  F02 also include age estimates
for each cluster in their sample based on the colour-magnitude
diagram morphology of the clusters.  We chose to adopt this
relatively homogeneous set of measurements as our reference
sample for open clusters.   The true values of age and metallicity
for many clusters are surely still systematically uncertain
by substantial amounts; we adopt the numbers from Friel and 
collaborators here because they place a large number of
clusters onto a consistent scale that is compatible
with the CG97 globular cluster scale (see below).

We picked specific
equatorial and southern clusters to observe based on richness
and coverage of the age-metallicity plane.  NGC 2682 (M67)
is a much-observed target, although it is not a convenient
one for FORS2 because of its extremely bright member stars
and large angular size.  However, it remains a useful calibrator
for metallicities close to solar.
Our open cluster RGB targets are listed
in Table \ref{open-tab}, with references to the photometry
and identifications.

\subsubsection{Star Cluster Metallicity Scales}

There is significant uncertainty in the absolute
metallicity scales of globular and open clusters.
Efforts to establish large samples of clusters
analyzed in a uniform way (consistent effective 
temperature scales,
model atmospheres, and lines analyzed) 
have not generally overlapped between the two 
cluster families.  It is therefore important to
establish that our adopted cluster metallicities
are not grossly incompatible with each other.

The F02 spectroscopic indices were calibrated to
a sample of field and cluster stars with high-dispersion
analyses.  Two globular clusters, M71 and M5, were
included; the reference metallicities were taken
to be $-$0.79 \citep{sne94} and $-$1.17 \citep{sne92}
respectively.  These are lower by 2$\sigma$ 
than the metallicities on the CG97
scale, $-$0.73 $\pm$0.03 and $-$1.12 $\pm$ 0.03 
(R97b).  Most of F02's
calibration sample were field giants measured by
\citet{cot86} and analyzed by \citet*{she93}. 
These authors used MARCS atmosphere models
\citep{gus75} and effective temperatures derived
from imposing excitation equilibrium of Fe~I and Ti~I
in the spectra.  In contrast, CG97 used stellar
atmospheres from \citet{kur93} and derived T$_{eff}$
from a V$-$K color-temperature relation.  

While the two halves of our reference sample are 
inhomogeneous, there is good reason to think they
are not inconsistent.  Inspection of abundances
derived from the same observational material
using MARCS versus Kurucz atmospheres \citep[e.g.][]{kra03}
indicates that the MARCS-based abundances tend to 
be lower by $\approx$ 0.08 dex.  Comparing the
differences due to T$_{eff}$ scales is more problematic,
given the extremely heterogeneous nature of the material.
We can estimate
the effects of the inconsistent T$_{eff}$ scales by
noting that: {\it a)} temperatures derived from the infrared
flux method \citep*[IRFM;][]{alo99} tend to be in excellent
agreement with those derived from Fe~I excitation
equilibria in MARCS models \citep[e.g.,][]{hil00},
and {\it b)} \citet{kra03} find that IRFM temperatures
tend to be $\sim$50~K lower than color temperatures,
resulting in abundance values higher by $\approx$ 0.05
dex when the former scale is used.  

There are two
implications of these trends: first, that the 
systematic offsets in abundance between the adopted
globular and open cluster metallicity scales are
comparable to or smaller than the internal uncertainties
within the respective scales; and second, that when
comparing the F02 metallicities to the CG97 metallicities,
the offset caused by choice of model atmospheres
may be of similar magnitude but opposite sign to the
offset caused by choice of T$_{eff}$ scales.  
However, offsets between the two metallicity scales
of order 0.1 dex are probable; the 
effects of adopting different scales are discussed
further in section \ref{sum-sec}.

\subsubsection{Trumpler 5}

In addition to the five well-studied open clusters, we observed
the important, but understudied, cluster Trumpler~5 
($\equiv$ Collinder~105 $\equiv$ C~0634+094) in order to provide
a test of our abundance calibration.   Tr~5 is among 
the most massive open clusters in the Galaxy 
\citep[][hereafter K98]{kal98},
but its high reddening has made it an unattractive object of
study for the most part.  Tr~5 is an interesting object 
because there is no published spectroscopic
abundance measurement, so we can immediately apply
our calibration to a derivation of its metallicity.
One photometric abundance estimate places it near solar
metallicity (K98), and it is of a richness
and distance that would make it a more convenient future calibrator
of the CaT in the moderate-abundance regime than M67 has been.
It will be critical to extend this calibration to solar
abundance and above as the CaT method is applied throughout
the giant and dwarf galaxies of the Local Group.  The 
metallicity of Tr~5 is in itself a valuable datum, 
because of the possible membership of the N-type 
carbon star V493~Mon in the cluster \citep*{kal74}.
An accurate measure of the cluster metallicity would
in turn lead to a more precise age estimate for the
cluster, and possibly to new constraints on
models of carbon star formation.  We selected candidate members 
of Tr~5 using 2MASS J and K$_{\mathrm S}$ photometry
\citep{skr97} and later made the cross-identification
with BV photometry of K98; targets are identified
in Table \ref{tr5-tab}.  Note that three stars appeared to
be differentially reddened in the (B$-$V, V) CMD, and that
lacking prior radial velocity information to guide selection,
two foreground stars were included in the sample.  V493~Mon
was observed in order to compare its radial velocity to
that of the cluster (no CaT abundance estimate can be 
made for C~stars).

\subsection{Data Acquisition and Analysis}

The observations were made in Visitor Mode at the 
Yepun (VLT-UT4) 8.2-metre telescope of ESO's Paranal Observatory,
with the FORS2 spectrograph in multiobject (MOS) mode, 
between 24 and 26 December 2002.  We used the 1028z$+$29
volume-phased holographic grism with the OG590$+$32 order
blocking filter. 
In MOS mode, the FORS2 field is covered by 
an assembly of 19 mechanical
slit jaws, 20$\arcsec$--22$\arcsec$ long, that can 
each be arbitrarily positioned along the horizontal
axis of the field (East--West for instrument rotator
angle = 0$^{\circ}$).  Throughout, we used a slit width
of 1$\arcsec$.  The red-optimized MIT/LL 2k$\times$2k
pair of CCDs were in use, with readnoise of 2.7 electrons
and gain of 0.8 $e^-$ ADU$^{-1}$.  In the standard
configuration, the CCD pixels are binned 2$\times$2,
giving a plate scale of 0.25$\arcsec$ per pixel
in the spatial direction.
This setup yields spectra covering
$\approx$ 1700 \AA, a central wavelength near 8500 \AA\,
and a dispersion of $\approx$ 0.85 \AA\ pix$^{-1}$ 
(resolution 2--3 \AA).
The FORS2 field is 6$\farcm$8 square,
but it is limited to 4$\farcm$8 useful width in the
dispersion direction in order to keep the wavelength
range of interest from falling off the edges of the CCD.
The typical exposure time per cluster was 3$\times$40
seconds, although the presence of 9th magnitude stars
in M67 led us to use exposures of just 3 seconds.
Skies were clear during the entire run; the 
seeing during these observations varied between
0$\farcs$5 $\la$ FWHM $\la$ 1$\farcs$2, with a
median stellar FWHM = 0$\farcs$6.

Astrometry of the targets was obtained via short
pre-imaging exposures with FORS2, taken several
weeks before the observing run.  The relative 
positions of the slit jaws were precisely set
before the observing run using the 
FORS Instrument Mask Simulator (FIMS) software
tool distributed by ESO.  Through-slit 
images were then taken immediately prior to 
the science exposures to confirm accurate slit
centreing, and the telescope was offset by up
to $\approx$ 0.25$\arcsec$ (1/4 of the slit width)
when necessary.  In cases where the targets
spanned a large range of $x$-values on the chip, 
it was noted at the time of observation that stars
at one end of the range could not be perfectly
centred simultaneously with stars on the opposite
end.  The differential offset was later estimated
to amount to roughly half a pixel over a 750 pixel baseline.
Implications of this are briefly discussed in the
following section.

Calibration exposures were taken during the 
daytime under the FORS2 Instrument Team's
standard calibration plan.  These consisted
of lamp flatfields with two different illumination
settings and a helium-neon-argon arc lamp exposure.
Two lamp settings are required for FORS2 flatfielding
to correct for extra reflections introduced into 
the optical path of the calibration assembly 
(T. Szeifert 2003, private communication).
Because of the large number of setups in our 
programme, twilight sky flats were deemed impractical.
The basic data reduction was performed in 
IRAF\footnote{IRAF is distributed by the 
National Optical Astronomy Observatories, which
are operated by the Association of Universities
for Research in Astronomy, Inc., under cooperative
agreement with the National Science Foundation.}.
We fit and subtracted the scaled overscan region,
trimmed the image, and divided by the appropriately
combined lamp flats within IRAF's {\tt ccdred} package.

The spectroscopic extractions were done using {\tt hydra},
an IRAF package for handling multislit spectra.
Our targets were bright
enough that the trace could be extracted directly
from the science exposures.  Across the field
of view, the curvature of the spectral trace varied
significantly, but could in all cases be well-fit
with a low-order polynomial.  Because of the 
potential for spectrograph flexure during the night
and the high probability of small, uncorrected slit-centreing
errors, we used night sky emission lines of OH 
\citep{ost92} and O$_2$ \citep{ost96}
to dispersion-correct the spectra instead of the daytime
arc lamp exposures.  Typically, 25--35 strong lines
were used in the wavelength solutions, which had
characteristic root-mean-square (r.m.s.)
scatter of 0.04--0.08 \AA.  The exception
to this rule was M67, for which the exposures were
so short that no sky lines were visible.  The 
spectral window covered depends on the position
of the slit along the dispersion direction of the 
CCD; some spectra reach as far into the blue as 7300 \AA, 
and some as far into the infrared as 10100 \AA, but 
most are centred near 8500 \AA\ and cover the 
approximate range 7600 \AA\ $\la$ $\lambda$ $\la$ 9350 \AA.

\begin{figure*}
 \includegraphics[scale=0.8]{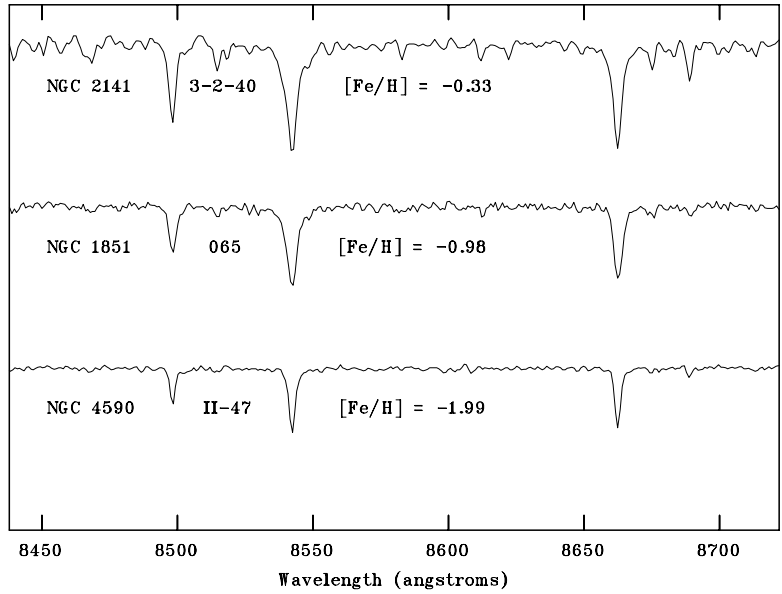}
 \caption{Sample spectra for stars in three clusters,
  representative of the range of line strengths
  in our targets.
  The spectra have been sky-subtracted,
  continuum-normalized, and Doppler-shifted to 
  the rest frame for comparison.  Each spectrum
  is offset by a constant amount from the one
  below.  Stars are identified by cluster name
  and metallicity from Table \ref{clus-tab},
  and ID from Tables \ref{glob-tab} and \ref{open-tab}.
  Most of the ``noise'' in the most metal-rich
  spectrum is due to the increasing strength
  of a large number of relatively weak atomic
  lines, mainly Fe~I.
  \label{spec-fig}}
\end{figure*}

Extraction to 1-dimensional spectra was done within
the {\tt apall} tasks.
Sky subtraction was performed
using 1-dimensional fits across the dispersion
direction.  Because the targets are bright compared
to the sky and the slitlets are long compared to the
seeing disc, this presented few difficulties, 
except when the stars fell near the ends of the slitlets.
In these cases, the sky region was chosen interactively
and adjusted to produce the cleanest extracted spectrum 
in the region around the CaT.  Dispersion-corrected
spectra were combined using {\tt scombine} to reduce
noise from cosmic rays and CCD defects.  The stars
were continuum normalized by fitting a polynomial to the 
spectrum, excluding the Ca II lines and regions of
strong telluric absorption.  We achieved typical signal-to-nosie
ratios of 20 $\la$ S/N $\la$ 85 per pixel.
Sample spectra showing
the region of the CaT appear in Figure \ref{spec-fig}.

\subsection{Radial Velocities}

Stellar radial velocities are important for deriving the
expected centroid of the Ca lines in the equivalent width
measurement, and in some cases for establishing cluster 
membership.  We are not primarily interested in the
absolute radial velocities of the target stars, so we
did not observe radial velocity standards.  Rather, we
cross-correlated our spectra \citep{ton79} 
with template stars already in our abundance programme.
This guarantees a good match in spectral type between
the templates and the stars with unknown velocities.
The templates were picked based on reliable velocity
measurements from the literature.
We chose measurements of 12 stars in NGC~4590, NGC~2298, and NGC~1904
\citep{gei95}, 10 stars in Berkeley~20 and
Berkeley~39 \citep{fri02}, and 2 stars in 
Melotte~66 \citep{fri93}.  
We used the IRAF {\tt fxcor} task
to perform the cross-correlation for each of the 24 templates,
and determined the stellar radial velocities using the
average of template offsets weighted by the error in the
correlation result and the height of the correlation peak.
The velocities were corrected to the heliocentric reference
frame within {\tt fxcor}, after suitable alteration
of the ESO image headers to the format demanded by the task.

When the stellar image is smaller than the slit width, the
random error in the velocities can be much smaller than 
the systematic error due to slight misalignments between
the slit centre and the stellar centroid \citep[e.g.,][]{irw02}.
When the seeing was better than 1$\arcsec$, we evaluated the
slit-centreing accuracy of each target by visual inspection
of the through-slit images.  Where necessary, we applied
corrections to the velocities of both templates and targets.
Our spectral resolution corresponds to $\approx$ 29.5
km sec$^{-1}$ pix$^{-1}$, yielding corrections that range from
$|\Delta$v$|$ = 0--32 km sec$^{-1}$.  In most cases, the 
offset between slit and star was strongly correlated with
the stellar $x$-position on the CCD.  
We estimate that the limit of 
centroiding accuracy is roughly a quarter of a pixel, 
or $\approx$ 7 km sec$^{-1}$.  We therefore adopt this value
as a floor in the error estimates on the heliocentric radial velocities
in Tables \ref{glob-tab}--\ref{tr5-tab}.  This is a 
conservative estimate, as the computed velocity dispersions
for some of the clusters are less than this floor.

The mean derived cluster velocities are given in 
Table \ref{vrad-tab}, with the r.m.s.\ dispersion and
a previous measurement from the literature, where
available.  In most cases, the cluster velocities
are in reasonable agreement with previously published
values.  For the globular clusters, there is a 
mean velocity shift of 4.7 km sec$^{-1}$ compared to
the literature velocities. 
Exposure times for M67 were so short that no 
sky lines could be measured for a good dispersion
solution, so we simply adopted the shape of the
function from other targets, and shifted the zeropoint
using the telluric A band of O$_2$.  Therefore 
we have no independent measure of V$_r$ for M67; radial velocities
from the list of \citet{mat86} are listed in 
Table \ref{open-tab}.  For comparison,
the mean V$_r$ from F02 is shown in Table \ref{vrad-tab}.

No measurement
of Trumpler~5's radial velocity could be found in the
literature; it will be discussed further in section \ref{tr5-sec}.
Of the four remaining open clusters, agreement with
previously published radial velocities is relatively
good for three: Be~20, Be~39, and Mel~66 are measured
within 10 km sec$^{-1}$ of the published values.  
However, NGC~2141 is found to be blueshifted more than
30 km sec$^{-1}$ from the velocity reported in \citet*{fri89}.
The source of the discrepancy is unknown, but our 
value is in good agreement with recent medium-dispersion
measurements independently made at the Lick Observatory
by one of us (TLB).  In all, five stars with velocities that 
were highly discrepant from their respective cluster means
were judged nonmembers and excluded from further 
analysis; they are flagged in the appropriate tables.

\subsection{Equivalent Widths}

We used interactive software kindly provided
by G. Da~Costa (1999, private communication to
TSH), and later modified by AAC,
to fit the line strengths of the Ca~{\small II}
lines.  The line and continuum bandpasses are 
as defined in \citet{arm88}, Doppler-shifted
in each case to the measured radial velocity
of the target.  The CaT lines can be contaminated
by weak neutral metal lines, and the continuum
may be affected by weak molecular bands.  Both
effects make it impossible to measure the ``true'' 
equivalent widths of the CaT lines in isolation.
The spectra need not be flux-calibrated because
the continuum slopes of red giants in this wavelength
range are generally close to flat, so a simple
normalization by an arbitrary polynomial suffices
for measurement of the so-called ``pseudoequivalent
width'' \citep{arm88}.
The measured line strengths depend on spectral
resolution, and measurements of a given star
by different groups often deviate from each other.
However, \citet{rut97a} and \citet{cen01a} have shown that
measurements of the CaT index on various systems
are generally readily transformable, so this is
not a worry.

\subsubsection{Fitting the Line Profiles}

Our initial approach, following \citet{arm91} and
\citet{sun93}, was to fit a Gaussian function to the
Ca~{\small II} lines.
However, it soon became apparent that for the stars
in the high-metallicity half of our sample, the line
profiles strongly deviate from Gaussian because of the
strong damping wings.  This effect
is even stronger for stars in our LMC sample (Cole et al.\
2003, in preparation), because of their lower average
surface gravities and hence broader lines.  Therefore
we modified the profile-fitting software to fit each
line with the sum of a Gaussian and a Lorentzian function:

\begin{eqnarray}
\label{fit-eqn}
\lefteqn{F(\lambda ) = C(\lambda )} \nonumber \\
 &  & - \alpha _1 \exp\left( -\frac{1}{2}\left( \frac{\lambda - \lambda _0}
  {\sigma}\right) ^2\right) 
  - \frac{\alpha _2 \Gamma}{\left( \lambda - \lambda _0\right) ^2
  + \left( \frac{\Gamma}{2}\right) ^2},
\end{eqnarray}

\noindent where the flux at wavelength $\lambda$ 
is equal to the continuum level C($\lambda$)
minus the summed
Gaussian and Lorentzian functions, with respective
amplitudes $\alpha _1$ and $\alpha _2$, and common
line centre $\lambda _0$.  The width parameters
$\sigma$ and $\Gamma$ have their usual
definitions.  Note that the continuum level is
effectively fixed at 1 by our normalization 
procedure.

The best-fitting parameters were determined in a 
least-squares way, using the Levenberg-Marquardt
algorithm as implemented by \citet{pre92}.  
Initial guesses were based on the line depth and
breadth in order to maximize speed of convergence.
There are no significant differences between 
the shape of the composite profile and the 
actual line shapes, a much more satisfactory
situation than the Gaussian fits.
The fits are then integrated
over the line bandpasses to give the equivalent
widths. 
Error estimates are obtained by measuring 
the root-mean-square scatter of the data about the
fit function.  

\begin{figure*}
 \includegraphics[scale=0.8]{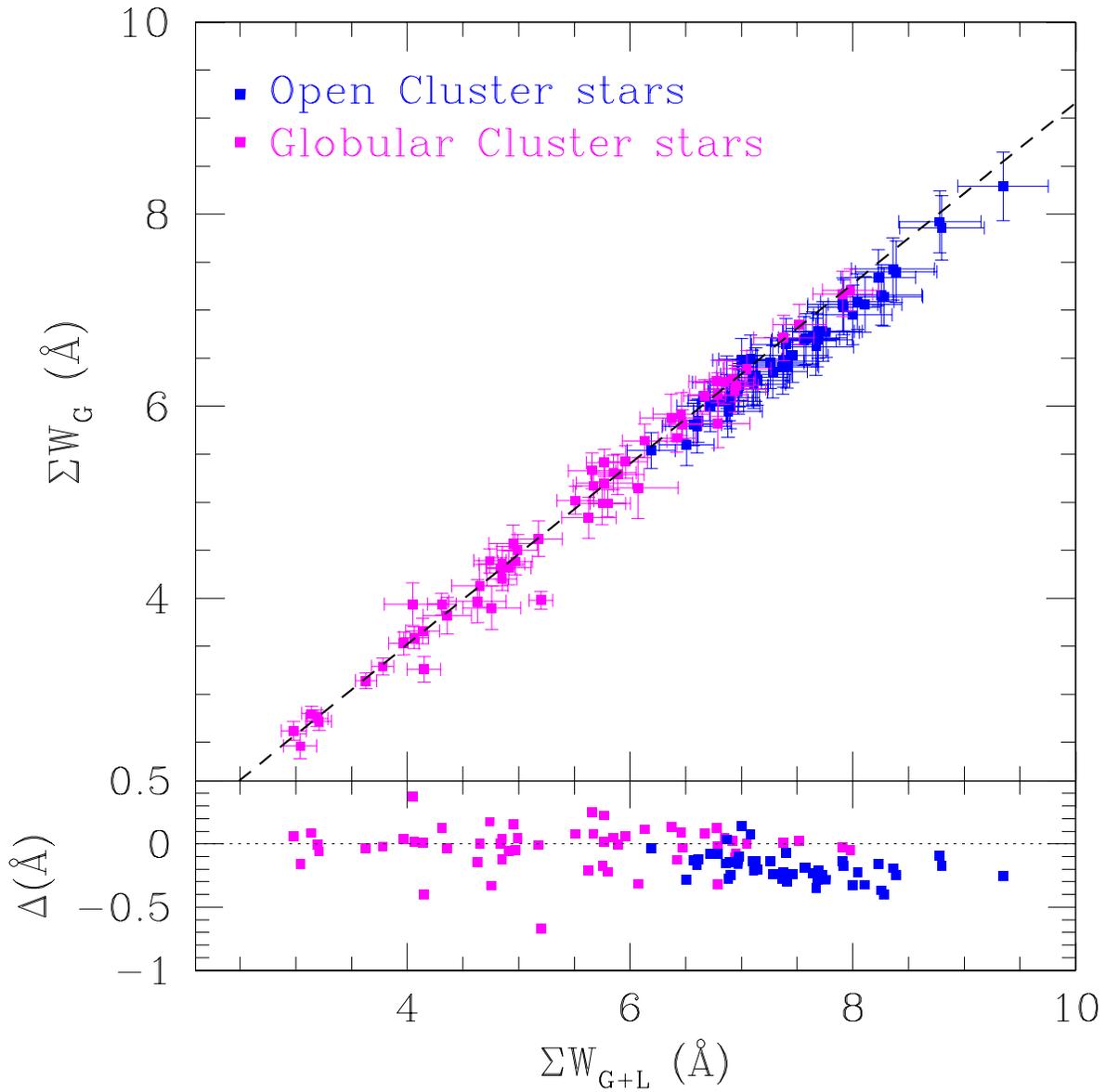}
 \caption{(Top panel) Summed equivalent widths of the 3 CaT 
  lines measured using a Gaussian fit, compared
  to the same lines fit with the sum of a Gaussian
  and a Lorentzian function.  The dashed line
  shows the best linear fit to the globular cluster
  stars. (Bottom panel) The residuals to the fit
  are plotted, showing how a Gaussian fit 
  increasingly fails to measure the
  rising fraction of absorption in the far line 
  wings as the lines get stronger.
  \label{comp-fig}}
\end{figure*}

The composite function fit provides a more 
satisfactory fit to the line shape, but 
does it change the resulting equivalent width
measurement?  We tested the effect of choice
of fitting function by remeasuring our targets
with the amplitude of the Lorentzian 
component equal to zero.  Figure \ref{comp-fig}
shows the comparison.
The relation between the Gaussian-fit
equivalent width, $\Sigma$W$_G$, and the composite-fit,
$\Sigma$W$_{G+L}$ is linear over much of its range,
but there is a change in slope at
$\Sigma$W$_{G+L}$ $\approx$ 6.5 \AA.  A least-squares
fit to the globular cluster stars produces the
relation

\begin{equation}
\Sigma W_G = 0.940\, \Sigma W_{G+L} - 0.241,
\end{equation}

\noindent with r.m.s. scatter of 0.11 \AA\ about 
the fit for the 57 globular cluster stars.  This
relation is plotted with a dashed line in the
upper panel of Fig.\ \ref{comp-fig}. 
The bottom panel of the figure shows the residuals
to this fit, demonstrating that the open cluster
stars have similar scatter to the globulars, but
are systematically offset by a mean residual
of $-$0.2 \AA\ ($\approx$ 2--3\%) from the fit.  This can be 
intuitively understood by noting that with 
increasing line strength, an increasingly larger
fraction of the absorption occurs in the line wings,
resulting in an ever-larger fraction of absorption
missed by the Gaussian fit.  The offset
corresponds to a difference of 0.05-0.15 dex in the
derived metallicities of strong-lined stars.

The ``gross underestimation'' of the strength
of the line wings by a Gaussian fit was already
noted by R97a.  
Because of the increasingly
important contribution of the damping wings to
the total line width for the strongest lines,
we therefore consider only the equivalent 
widths derived from composite line fits
in the remainder of this paper.  In the 
next section, we show that with this choice,
the CaT index maintains sensitivity over the
range of metallicities considered here.  
However, the composite fit is noisier than
the Gaussian fit because the line depth is
small in the far wings.  By adding artificial
noise to one of our spectra and remeasuring
the equivalent widths, we estimate that
for signal-to-nosie ratios less than $\approx$ 15--20,
there is no significant difference between
the two profile shapes.  Thus where spectra
of low S/N and/or of low metallicity
stars are concerned, a fit with fewer free
parameters, such as a Moffat function (R97a)
may be preferable.

\subsubsection{Forming the CaT Index}

Metallicity measurements using 
the CaT index use a linear combination of the
three individual Ca~{\small II} lines to form
the line strength index $\Sigma$W \citep{arm91}.
Various authors have used different procedures
to form this index, depending on spectral resolution
and signal-to-noise.  Some authors have excluded
the weakest ($\lambda$ =  8498 \AA ) line
\citep[e.g.,][]{sun93,col00}.  Others
have used all three lines, either weighted
\citep[R97a,][]{sme03} or unweighted \citep{ols91}.
Because of our extremely high signal-to-noise ratios,
we use an unweighted sum of the three lines.  The
fractional errors on the individal lines are roughly
equal, and variations within the sample make it
inadvisable to apply a uniform weighting scheme.
uniform weighting scheme for all clusters.  It has
been shown by R97b and \citet{sme03} that the various means
of creating $\Sigma$W produce well-defined linear
mappings between the resulting index values.
For the remainder of this paper, we adopt the
convention

\begin{equation}
\Sigma W \equiv W_{8498} + W_{8542} + W_{8662}.
\end{equation}

Stellar effective temperature and surface
gravity play a strong role in forming
the Ca~{\small II} lines in red giants,
and so the metallicity dependence is typically
found using the so-called ``reduced'' equivalent
width \citep{arm91}.  This quantity is defined
as 

\begin{equation}
\label{wpr-eqn}
W^{\prime} \equiv \Sigma W +
  \beta\, (\mathrm{V-V_{HB}}),
\end{equation}

\noindent where the definition exploits
the empirical fact that in an individual
globular cluster, all red giants at a given
V magnitude have the same $\Sigma$W, and 
that $\Sigma$W (V) increases linearly
up the RGB. 
V$_{\mathrm{HB}}$ here is
the mean magnitude of the horizontal branch;
its introduction removes any dependence on
the cluster distance or reddening.  Other
parameterizations for W$^{\prime}$ are 
possible \citep[e.g.,][]{ols91}, but 
Equation \ref{wpr-eqn} has the advantages
that a huge amount of V band photometry
is available in the literature, and 
that no additional color-dependence is introduced. 
R97a calibrated Equation \ref{wpr-eqn} using
data for 52 Galactic globular clusters
and derived a slope $\beta$ = 0.64 $\pm$0.02
\AA\ mag$^{-1}$, independently of metallicity,
across the range $-$2.1 $\la$ [Fe/H] $\la$ $-$0.6.
Consistency with this constant $\beta$ has been
found in smaller studies including more metal-rich
(open) clusters, such as 
Melotte~66 \citep{ols91}, 
NGC~2477 \citep{gei95}, and M67 \citep{ols91,col00,sme03}.
Defined as in Equation \ref{wpr-eqn},
every cluster can be assigned a single value
of W$^{\prime}$ that is strongly correlated with
metallicity.  The relation is linear for the
CG97 metallicity scale, but mildly nonlinear
for other choices (R97b).

Because we use a different functional form
to fit the Ca~{\small II} lines of our stars,
we also need to rederive $\beta$.   We did 
so by deriving the slope for each cluster
and taking the average, weighted by the
error in the individual values.  The resulting
slope is $\beta$ = 0.73 $\pm$0.04.  This 
value is higher than found for the globular
cluster-only sample of R97a.
For reference, the slope determined from 
our sample of five globulars alone is 
$\beta _{glob}$ = 0.66 $\pm$0.03, which 
is indistinguishable from the favored value
of R97a, but $<$2$\sigma$ different
from our value for all 10 clusters.
From the models of \citet{jor92}, it can
be inferred that an increase in $\beta$ 
with metallicity is in qualitative agreement
with theory; direct comparison is difficult
because of the interplay between abundance,
gravity and temperature and how these
quantities vary with age among RGB stars.

\begin{figure*}
 \includegraphics[scale=0.8]{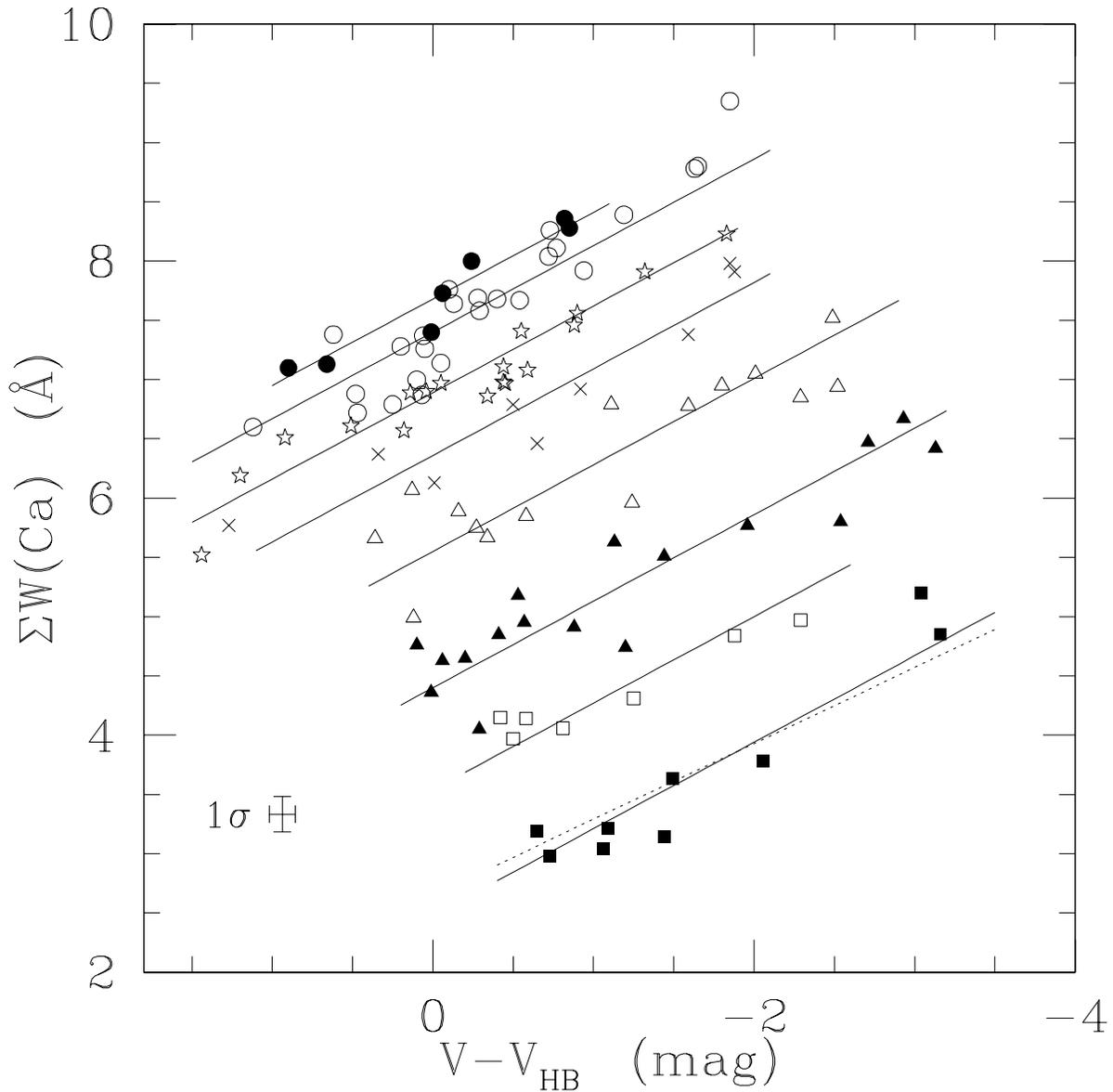}
 \caption{The CaT equivalent widths of RGB stars in 
  10 star clusters compared to their magnitude above
  their respective cluster horizontal branches.
  The solid lines show the best-fitting lines to each
  cluster, assuming the best common slope of 
  0.73 \AA\ mag$^{-1}$.  
  Clusters are represented
  by: closed squares (NGC~4590), open squares (NGC~2298),
  closed triangles (NGC~1904), open triangles (NGC~1851),
  crosses (47~Tuc), stars (Mel~66 and Be~20), open circles
  (NGC~2141 and Be~39), and closed circles (M67).
  The fit to NGC~4590 adopting the slope from R97a of
  0.64 \AA\ mag$^{-1}$ is shown for comparison (dotted line).
  Errorbars are omitted for clarity, but the typical 1$\sigma$
  errors are shown at lower left.
  \label{vhb-fig}}
\end{figure*}

The equivalent width measurements for
the individual stars are shown in Figure
\ref{vhb-fig}, with photometry drawn from the 
sources listed in Tables \ref{glob-tab} \&
\ref{open-tab}.  Error bars are omitted
so that the best-fitting line (with slope = 0.73) 
to each cluster can be clearly seen.  Note
that there is significant scatter about the 
cluster fiducials even for the well-studied
globulars.  The mean r.m.s. scatter about
each fiducial is $\pm$0.24 \AA, 
larger than our estimated measurement errors;
the same effect was found by R97a in their
sample.
Note that some cluster pairs with 
metallicities that differ by  
$\sim$0.1 dex (Melotte~66 and Be~20; 
NGC~2141 and Be~39) have been plotted using 
a single point type and fiducial line for 
clarity.  Clusters of such similar abundance
are at the limit of our metallicity
resolution.

\section{Calibrating the Metallicity Scale}
\label{cal-sec}

Most previous work has found a linear 
relationship between the reduced equivalent
width and [Fe/H] 
\citep{arm91,ols91,sun93,dac95,gei95,rut97b,col00,tol01,sme03,kra03}.
The recent standard has been to calibrate to the
relation between W$^{\prime}$ and [Fe/H] on the
CG97 scale, as given by R97b.
Given that we have derived W$^{\prime}$ on a
totally different scale from R97b, we will 
need to rederive this relation. 

\begin{figure*}
 \includegraphics[scale=0.8]{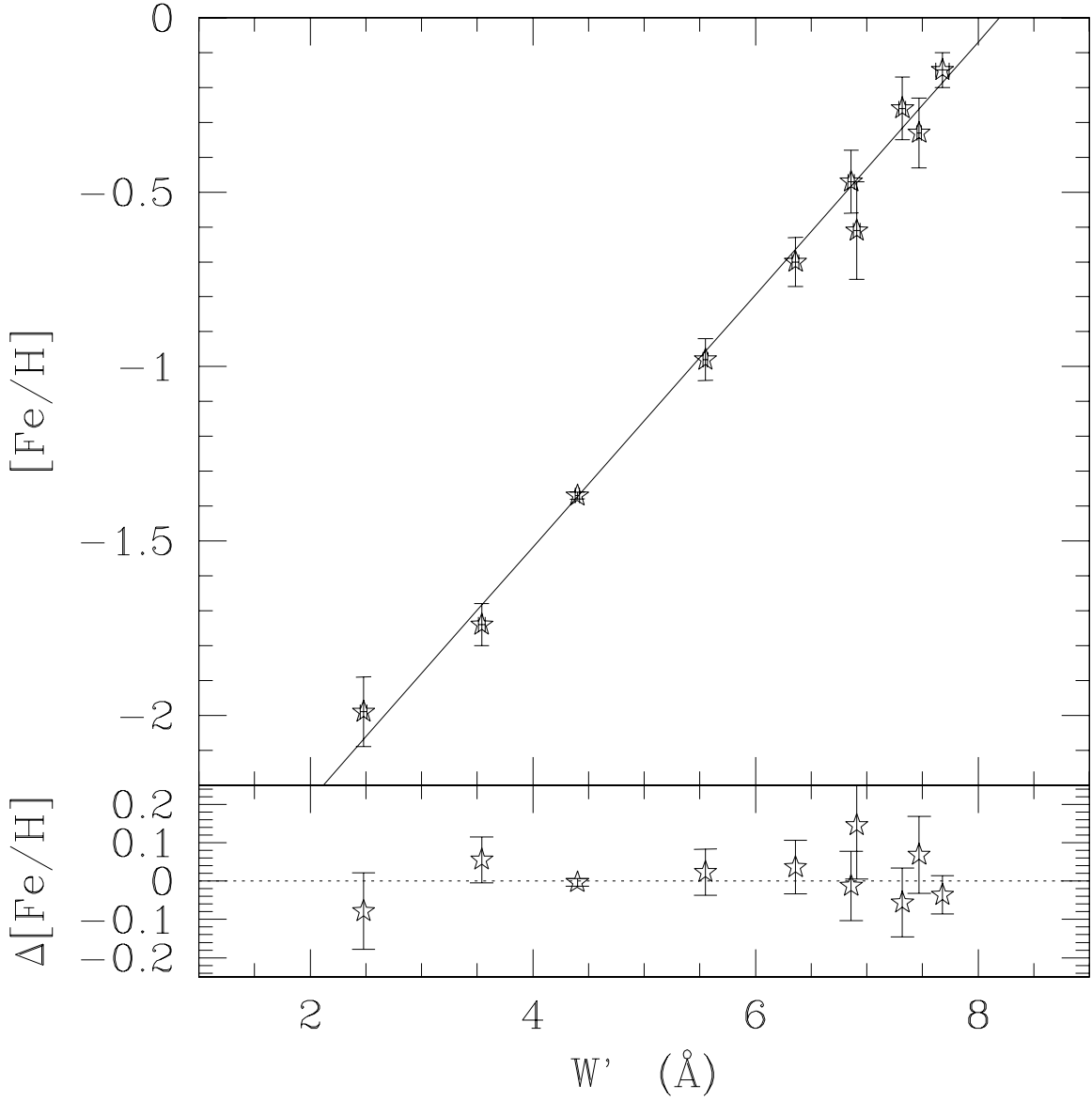}
 \caption{(Top panel) The reference metallicity of the 10 
  clusters in Table \ref{clus-tab} plotted against
  their reduced CaT equivalent width.  The best
  linear fit to the data is shown.  (Bottom panel)
  the residuals to the linear fit are plotted; the
  errorbars are 1$\sigma$ errors on the reference
  abundance.
  \label{cal-fig}}
\end{figure*}

Figure \ref{cal-fig} shows the cluster 
reduced equivalent widths plotted against
the reference abundance on the CG97/F02
scale.  The best-fitting line through the
data is shown, with the residuals to the
fit shown in the bottom panel.  The 
equation of the fit is

\begin{equation}
\label{cal-eqn}
\mathrm{[Fe/H]^{F02}_{CG97}} = (-2.966 \pm0.032)
       + (0.362 \pm0.014)\, \mathrm{W^{\prime}},
\end{equation}

\noindent 
with r.m.s.\ scatter 
$\sigma$ = 0.07~dex.  With this set of
reference abundances, there is no obvious
age effect on the calibration, nor is there
any dramatic deviation from the straight-line
fit for metallicities beyond the range 
of the R97b calibration.  The only cluster
that deviates by as much as 1$\sigma$ from
the fit is Be~39; we will discuss the
effects of alternate metallicity scales
in section \ref{sum-sec} below.

Some curvature may be present,
based on a predicted loss of CaT index
sensitivity at high metallicity
\citep*{dia89}.  Evidence
for just such an effect was presented by
\citet{car01}, who were driven
primarily by upward revisions of the
abundance of the metal-rich bulge globulars
NGC~6528 and NGC~6553.  If we follow
\citet{car01} in adding a W$^{\prime 2}$
term to equation \ref{cal-eqn}, we find
the coefficient of the quadratic term
to be formally insignificant, and the 
scatter is not measurably reduced.
We conclude that a linear relation suffices
to reliably predict [Fe/H] on the currently
adopted scale from our measurements of W$^{\prime}$.
However, there are two caveats:  

\begin{enumerate}
\item Homogeneous
high-resolution abundances for the open clusters
are not yet available, so upward revision of their
reference metallicities could change this conclusion.

\item Age-metallicity effects could be conspiring
to produce the observed linear relation, i.e., 
perhaps surface gravity and/or temperature effects
cause a variation in W$^{\prime}$ with age that
we do not observe because we lack the dynamic range
in age at high metallicity to do so.
\end{enumerate}

Both of these points are further discussed below.

\section{The Open Cluster Trumpler 5}
\label{tr5-sec}

Trumpler~5 was discovered nearly 75 years 
go by visual inspection of photographic
plates taken by E.E. Barnard in 1894 
\citep{tru30}.  The cluster lies in the
Galactic plane $\approx$ 20$^{\circ}$ from
the anticentre, in a region of high and
variable interstellar reddening that 
has complicated attempts to derive its
fundamental parameters.
Various CMD studies have
been undertaken, focusing on the red giant
branch in the 1970's \citep[e.g.,][]{kal74},
and reaching the unevolved main-sequence
in the 1990's \citep[K98;][]{kim03}.  Authors
have converged on values near 3~kpc for 
the distance, E(B$-$V) = 0.6 for the reddening,
and M$_{\mathrm V}$ = $-$5.8.   K98 recognized
that this makes it one of the most massive
known open clusters in the Milky Way; we
estimate it to be some 5 times as massive
as M67, and within a factor of 10 of the 
massive, intermediate-age clusters of the 
Magellanic Clouds. 

Tr~5 can thus be an important cluster
for the study of intermediate-age stellar
populations, cluster evolution, and population
gradients in the Galaxy.   However, no 
reliable metallicity measurement has yet been
made.  \citet{kim03} estimate [Fe/H] = $-$0.3
based on the photometry of K98; K98 simply
took the metallicity to be equal to that of 
M67.  Because of age-metallicity degeneracy,
the age is similarly uncertain; the problem
is worsened by the effects of reddening.
From the same photometry, K98 and \citet{kim03}
derive 4.1~Gyr and 2.4~Gyr, respectively,
while the morphological age index of \citet{jan94}
yields 4.9~Gyr.  

We included Tr~5 in our observing plans for
three reasons: 1) to obtain the first accurate
metallicity assessment for this cluster; 2) to
make a testable prediction on our calibration
of the Ca~{\small II} triplet method at high metallicities
and intermediate ages; and 3) to lay the groundwork
for subsequent use of red giants in Tr~5 as 
metallicity standards for application of the
CaT method from 8--10~metre class telescopes
(for which nearby clusters are inconveniently bright).

The observed stellar sample towards Tr~5 is 
given in Table \ref{tr5-tab}, with the adopted
V magnitudes, the derived radial velocities,
and equivalent widths.  No radial velocities
have previously been published, so we were 
unsure whether or not we could reliably discriminate
between cluster members and the field.  However,
we found a well-determined mean heliocentric radial
velocity V$_r$ = $+$54 $\pm$5 km sec$^{-1}$, where
the uncertainty reflects the r.m.s. scatter of the
member stars about the mean.  This squarely assigns
Tr~5 to the Galactic disc, but it may be lagging
in rotation
by 10--20 km sec$^{-1}$ compared to the thin disc-- 
which is unsurprising, given its estimated age.

Two stars were 
excluded from further analysis based on their blueshifts
of 40 and 60 km sec$^{-1}$ respectively relative to
the cluster-- consistent with them being dwarfs 
in the solar neighborhood.
We initially chose our stars based
on the 2MASS (K$_{\mathrm{S}}$,
J$-$K$_{\mathrm{S}}$) CMD, but later
inspection of the K98 (V, B$-$V) data showed that
three of the stars in the southern part of the
cluster appeared to be differentially reddened 
away from the main cluster RGB; these stars are
identified and excluded from further analysis.
Two of the stars near the observed tip of the 
red giant branch were treated with care because
the cluster is known to harbor M~giants, for which
TiO veiling can bias the CaT-derived metallicities
\citep{ols91}.
However, any TiO bands must be weak, or they 
would have been obvious in our spectra.
We decided to include
the 2 possible early M~stars because they don't 
significantly change the cluster metallicity, so 
any continuum veiling bias must be small.

\begin{figure*}
 \includegraphics[scale=0.8]{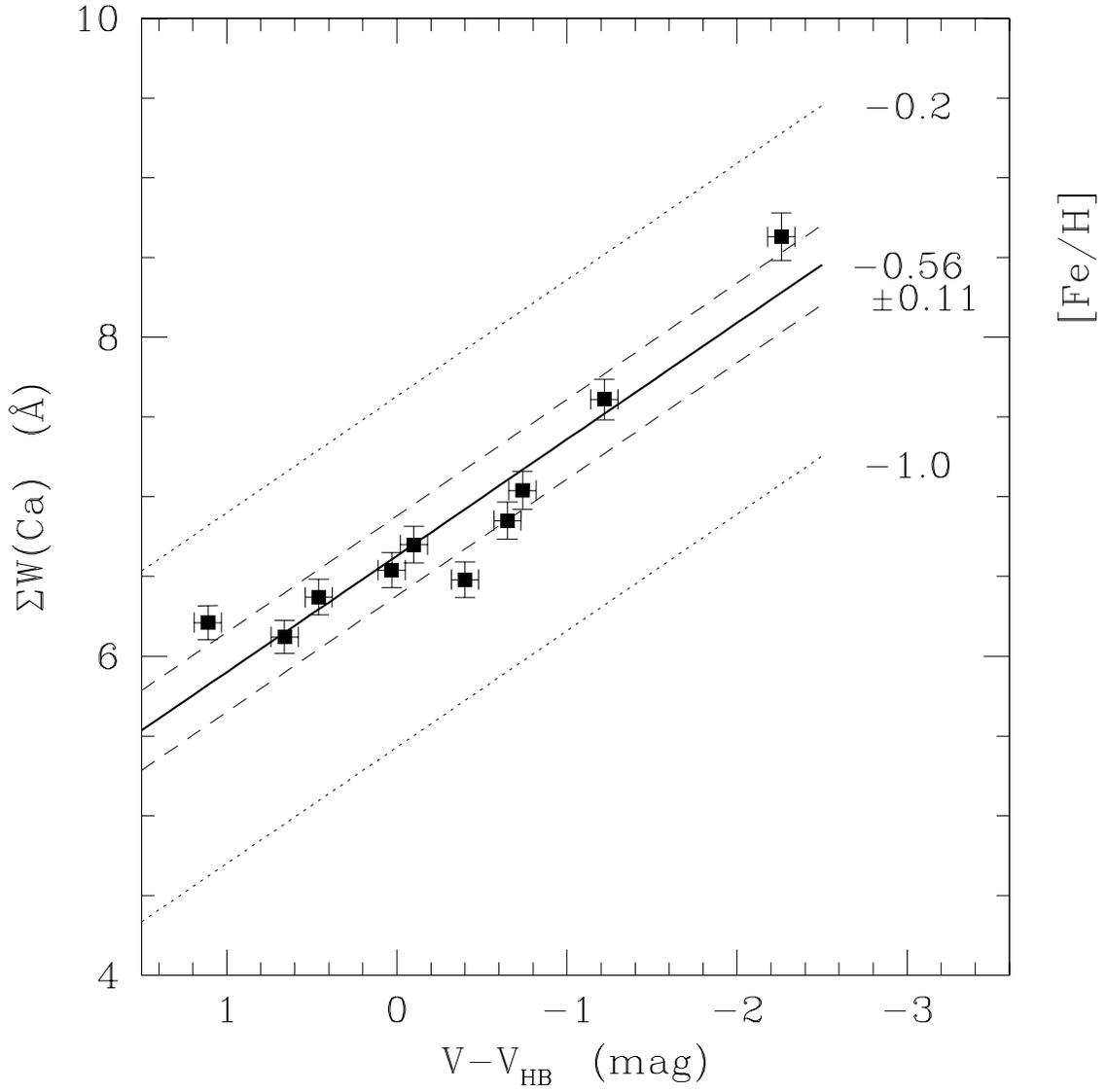}
 \caption{CaT equivalent widths vs.\ V magnitude
  for 10 RGB stars in the old open cluster
  Trumpler 5.  The lines corresponding to the 
  best-fitting metallicity according to our calibration
  (solid), and the formal 1$\sigma$ rms scatter
  (dashed) are shown, with reference lines (dotted)
  corresponding to $-$0.2 and $-$1.0 dex.
  \label{tr5-fig}}
\end{figure*}

The equivalent widths and magnitudes of the ten 
remaining RGB stars are shown in Figure \ref{tr5-fig}.
From these measurements and Equation \ref{cal-eqn}
we derive the metallicity
of Trumpler~5 [Fe/H] = $-$0.56 $\pm$0.11.  Lines
corresponding to this value and range are 
shown in the figure.  This is lower than the
metallicities estimated from the K98 CMDs,
but consistent with its large galactocentric
radius ($\approx$ 11~kpc). 
A full re-analysis of the CMDs is
beyond the scope of this paper, but we 
use isochrones from \citet{gir00} 
interpolated to the correct metallicity
to rederive the cluster age.  Using the
VI data, we find good
agreement with the main-sequence, subgiant
branch, and lower RGB for a distance of 
2.8~kpc, a reddening E(B$-$V) = 0.66 mag,
and an age of 3.0 $\pm$0.5~Gyr.  The age is
between the values reported by K98 and \citet{kim03},
for 10--14\% higher reddening and 7--18\% smaller
distance.

\citet{kal74} observed
that the carbon star V493~Mon is projected
within 2$\farcm$5 of the cluster centre and
has typical optical magnitudes for an N-type C~star
if assumed to lie at the cluster distance
and reddening.  This is also true if one considers
the near-infrared photometry from \citet{iij78}.
Assigning a membership probability will require precise
radial velocities and proper motions; 
however, a highly discrepant radial velocity
is sufficient to rule out membership.  We find
V$_r$ = $+$46 km sec$^{-1}$, close to the 
cluster mean.  Considering its projected
distance within 1/4 of the cluster radius of the
centre, the surface density
of field carbon stars, the optical photometry,
the infrared photometry, and the radial velocity,
V493~Mon seems to be a more than likely member
of Trumpler~5.  

\section{Application to Composite Populations}
\label{mix-sec}

The CaT method is the best available route
to metallicity measurements of RGB stars
in nearby galaxies \citep[e.g.,][]{sun93,tol01}.
There are two potential difficulties
in its application.  First, the 
reduced equivalent width, and its correlation
with [Fe/H], are only defined and calibrated
to stars older than $\sim$10~Gyr and more
metal-poor than $\sim \frac{1}{3}$ solar.
While this does not present severe problems
for some of the smallest dwarf spheroidals
(Draco, Ursa~Minor, Tucana), most Local
Group galaxies contain substantial populations
that violate either the age or metallicity
range of the R97b calibration \citep{mat98}.
We have 
begun to address this calibration issue
with these data.  

Second,
star clusters give a strongly biased view
of the star-formation history of a galaxy,
so the field stars are of primary interest.
Yet in a composite sample, the unknown 
mixture of ages and metallicities makes
it impossible to assign a unique horizontal
branch magnitude that is to be used in the
abundance determination of an individual star.
The observed horizontal branch (or more generally,
red clump) is more extended in the field than 
the cluster, because of the variation of clump
magnitude with age and metallicity
\citep[e.g.,][]{col98,gir01}.

A first approach to this problem was made
by \citet{dac98} in their study of SMC clusters.
These authors measured the mean red clump
magnitude in each cluster to create their
reduced equivalent widths, and then applied
a theoretically-derived correction factor to
the metallicity, 
based on the variation of clump magnitude
with age.  The corrections were of order $-$0.05 dex for
the age range from 3.5--10~Gyr, smaller than the precision
of the abundances.  We dispense with this type of
correction here, because
we have included younger clusters in our calibrations,
and the variation of their red clump magnitude 
is thus folded into Equation \ref{cal-eqn}.

\citet{col00} estimated the bias introduced
by the use of a composite V$_{\mathrm{HB}}$
in the calculation of W$^{\prime}$ for field
stars, and found similar results to \citet{dac98}.
We make a new, empirical measurement
of the bias without recourse to theoretical
models by considering all of our star clusters
together (``mixed''), rather than individually
(``unmixed'').  We do this
by shifting each cluster to a common apparent
distance modulus (m$-$M)$_{\mathrm V}$, and 
calculating the mean horizontal branch magnitude
of the mixed population, 
$\langle {\mathrm M_{V}}\rangle = 0.63$, with an
r.m.s. scatter of 0.14 mag.  Using new
values of V$-$V$_{\mathrm{HB}}$ calculated with
this mean HB magnitude, we recomputed the
metallicities of each cluster.  

\begin{figure*}
 \includegraphics[scale=0.8]{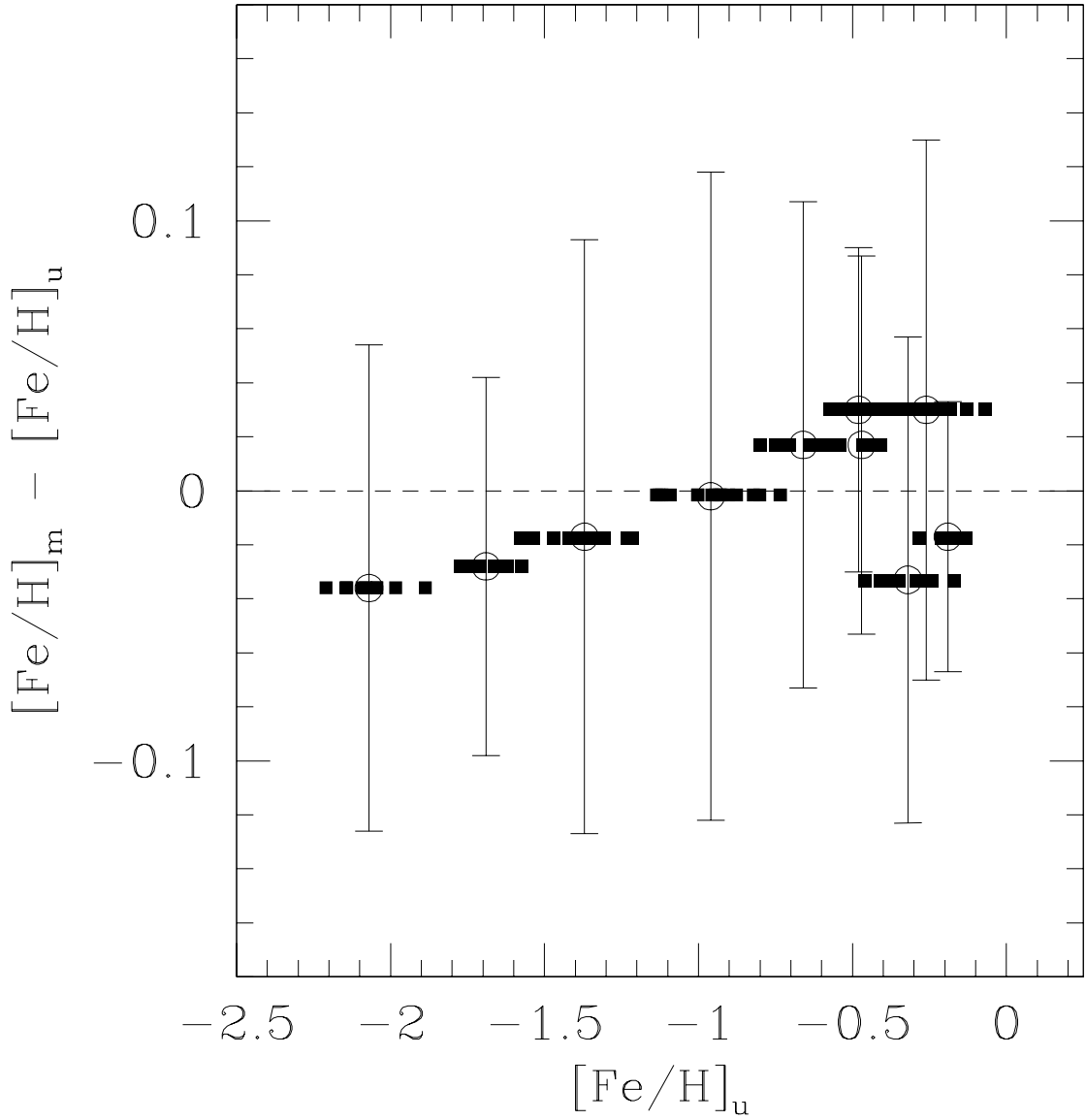}
 \caption{The difference between ``mixed''
  and ``unmixed'' abundance estimates for
  the 10 well-studied star clusters.  The
  ``mixed'' estimates are made by shifting
  all clusters to a common (m$-$M)$_V$ and
  using the combined average red clump $+$
  horizontal branch magnitude to define 
  V$_{\mathrm{HB}}$.  The errorbars show
  the rms scatter of the ``unmixed'' abundances
  of each cluster.
  \label{mix-fig}}
\end{figure*}

The result
of this exercise is shown in Figure \ref{mix-fig}.
The difference between the mixed vs.\ the 
unmixed metallicities is plotted against
the unmixed metallicities for the individual
stars in each cluster.  Since the difference
is constant within a cluster, the clusters spread
out along horizontal lines in this plot.
The errorbars in the $y$-coordinate give the
r.m.s. spread in metallicity of each cluster.
Figure \ref{mix-fig} clearly shows that the
bias introduced by variation of horizontal branch/red clump
magnitude with age is smaller than the intrinsic
precision of the method.  At low abundances,
a trend with metallicity is apparent, but at
the high end of the range, the effect is not
clear, perhaps due to errors in the distances
of the clusters, or perhaps due to a complicated
interaction between age and metallicity effects.
Figure \ref{mix-fig} is empirically supportive
of the theoretically-motivated corrections
suggested by \citet{dac98} and \citet{col00}.
The small size of the offsets indicates that
application of the CaT method to field populations
of mixed age and metallicity does not introduce
large biases in the derived metallicities, within
the range of parameters considered here.

\section{Summary \& Discussion}
\label{sum-sec}

In connexion with a programme to determine
the age-metallicity relation and radial
abundance gradient of field red giants
in the LMC \citep[][Cole et al.\ 2003
in preparation]{col00,sme03}, we have 
obtained medium-resolution spectra 
of red giants in 11 Galactic star clusters 
at the near-infrared Ca {\small II} triplet
(CaT)
using the FORS2 multiobject spectrograph
at the 8.2-metre VLT/UT4.  Our spectra
have extremely high signal-to-noise,
ranging from 20 $\leq$ S/N $\leq$ 85. 
Following many previous authors (e.g., R97b)
we have derived a linear relation between
the equivalent width of the CaT and the 
cluster metallicities.

For the first time, our sample includes
open clusters with a range of metallicities
from $-$0.6 $\la$ [Fe/H] $\la$ $-$0.2 and
ages from 2.5 $\la$ (age/Gyr) $\la$ 7. 
This allows an initial assessment of the
effects of age on the metallicity measurement
via the CaT.  While there is is substantial
scatter and the sample size is small, we
find no evidence for a decrease in sensitivity
of the CaT at young ages and high metallicities.
This behavior is theoretically supported by
non-LTE stellar atmosphere calculations
\citep[e.g.,][]{jor92} and is in good agreement
with comprehensive empirical calibrations of
the CaT behavior over a wide range of stellar
atmospheric parameters \citep[e.g.,][]{cen02}.
However, our results disagree with earlier 
results based on more limited stellar samples
\citep*[e.g.,][]{jon84,car86,dia89}.  The 
fact that we see no loss of metallicity 
sensitivity above [Fe/H] $\approx$ $-$0.5
is interesting in light of the observed 
metallicity independence of calcium line
strength in the {\it integrated} light of
clusters and galaxies above this level
\citep[e.g.,][]{vaz03}.

One difference between
our analysis and earlier work is that we fit
a composite function-- the sum of a Gaussian
and a Lorentzian-- to the line profiles.  This
is mandated by the line shapes, which indicate
that for abundance [Fe/H] $\ga$ $-$0.6, the
damping wings of the lines are so strong and broad
that we begin to resolve the line even at 
our resolution of $\sim$2.5 \AA.

Using our data, we obtain a linear relation between
the reduced equivalent width of a cluster, W$^{\prime}$,
and metallicity on a scale formed by the high-dispersion
results of CG97 for the globular clusters, and 
medium-dispersion Fe indices from F02 for the open clusters.
There is no indication that a second-order term in 
W$^{\prime}$, as advocated by \citet{car01}, is needed
to fit the data.  However, our sample does not include
old, high-metallicity clusters (such as bulge globulars
NGC~6528 and NGC~6553), clusters of supersolar metallicity
(such as NGC~6791 and NGC~6253), or young clusters of
very low abundance (such as the clusters of the Small
Magellanic Cloud).  Until such time as the age-metallicity
plane is filled, we cannot definitively rule out an
age effect on the calibration.  Filling in the high-metallicity
regime at a large range of ages will have important 
consequences for the determination of abundances in the
M31 halo.

An important point is that the globular cluster 
metallicity scale is not absolutely established.
The CG97 scale differs in a nonlinear way from the
average of spectrophotometric indices compiled by
\citet{zin84}.  In turn, CG97 may be superseded
by abundance scales derived from Fe~{\small II}
lines, rather than Fe~{\small I} lines, because
of the newly-recognized importance of non-LTE
effects \citep{kra03}.  The open cluster abundances
are even more uncertain, and will remain so until
large-scale, homogeneous surveys for high-dispersion
spectra can be undertaken.  This type of analysis
is presently available for only a small number 
of stars in a small number of clusters; among
our sample, Melotte~66 and M67 have been analyzed.
Photometric metallicity indicators based on 
DDO-system or Str\"{o}mgren photometry 
\citep[e.g.,][]{twa97} often give quite different
results from the indices tabulated by Friel and
collaborators.  The choice of metallicity scales
will of course have a strong effect on the
calibration of the CaT.  We tabulate some
alternate cluster metallicities in Table \ref{met-tab}.
Columns 1--4 give the cluster name, W$^{\prime}$
derived from Equation \ref{wpr-eqn}, the reference
abundance from Table \ref{clus-tab}, and the
derived metallicity using Equation \ref{cal-eqn}.
Alternate metallicities are given in column 5, sorted
by origin, and referenced in the last column.  Using
the tabulated values of W$^{\prime}$ and 
[Fe/H]$_{\mathrm{alt}}$, we derive the following
relation (excluding Tr~5):

\begin{equation}
\label{alt-cal}
\mathrm{[Fe/H]_{alt}} = -2.629\, + 0.058\, 
  \mathrm{W^{\prime}}\, + 0.037\, \mathrm{W^{\prime 2}}.
\end{equation}

\noindent  Significant curvature is introduced 
into the relationship because the globular cluster
metallicities have been shifted down, while the
open cluster metallicities, especially NGC~2141 and
M67, have been shifted upwards.  The cluster 
metallicities derived from Equation \ref{alt-cal}
are given in Table \ref{met-tab}.  We cannot 
comment on the correct choice of metallicity
scales, but we note that although the linearity
of the CaT calibration seems to be a fortuitous
result of one such choice, the one-to-one
correspondance between W$^{\prime}$ and
[Fe/H] holds no matter the choice of
reference metallicities \citep[as already noted
by][]{kra03}.

We have measured the bias introduced by 
taking the mean horizontal branch/red clump
magnitude of a composite field population
in determining the quantity V$-$V$_{\mathrm{HB}}$.
Shifting all 10 clusters to a common apparent
distance modulus and taking the mean V$_{\mathrm{HB}}$
of the mixed population results in derived
metallicities that shift by $\pm$0.05 dex
relative to the unmixed results.  This is
negligible within the uncertainties, and
shows that measurements of individual field
red giants yield reliable metallicities 
despite the lack of a unique V$_{\mathrm{HB}}$
in that scenario-- although the additional
uncertainty should be considered in the
error analyses.  The mixed-population bias
is clearly not random at the low-metallicity
end of our calibration sample, where the
age range is small.  This is consistent 
with theoretical expectations for the
variation of V$_{\mathrm{HB}}$ with
metallicity \citep[e.g.,][]{gir01}.
At the high-metallicity end, the picture
becomes more complicated, with age and
abundance effects combining with distance
uncertainties to introduce additional
scatter.

We have made the first measurement of the metallicity
of the large, old open cluster Trumpler~5, finding
[Fe/H] = $-$0.56 $\pm$0.11.  While interesting 
in itself, this also represents a testable
prediction of our calibration of the CaT.
With our metallicity estimate, the most likely
age for Tr~5 is 
3.0 $\pm$0.5~Gyr.  The relatively low metallicity of Tr~5
places the cluster in a category with
objects such as Melotte~66.
It somewhat diminishes
the utility of Tr~5 as an intermediate-age,
high-metallicity calibrator for the Ca~{\small II}
triplet, but makes it easier to theoretically
understand the possible presence of a
carbon star in the cluster.
The carbon star, V493~Mon, has a radial
velocity within 8 km sec$^{-1}$ of the
cluster mean.  Given the large number of
independent data consistent with membership,
it would be surprising if V493~Mon turned
out to be a field interloper.  However,
more precise velocities, as well
as proper motions, are needed to 
assign V493~Mon to Tr~5 for certain.
This raises the exciting possibility
that detailed study of the cluster
will yield its turnoff mass
($\approx$ 1.14$^{+0.09}_{-0.04}$ M$_{\sun}$),
and therefore
a direct measurement of the mass of 
the progenitor of an intermediate-mass,
moderate-metallicity carbon star.

We are further refining the empirical
calibration of the CaT by enlarging
the sample of high-metallicity,
intermediate-age cluster giants with high
signal-to-noise spectra.  TLB and TSH
have measured the spectra of numerous
giants in several clusters, at both 
high- and low-dispersion, using the
Shane 3-metre Telescope at Lick Observatory.
The cluster sample spans the abundance
range from $-$2.3 $\le$ [Fe/H] $\le$ $+$0.4
and ages $\ga$~2 Gyr.  In addition 
to the new CaT measurements, we will
determine [Ca/H] abundances for the
clusters using model atmosphere analyses
of the strengths of the plentiful neutral
calcium lines (which are theoretically much
better understood) in the high-dispersion
spectra.  We will calibrate the relationship
between the CaT index and [Ca/H], eliminating
the need to assume a specific [Ca/Fe] vs.\ 
[Fe/H] relationship.  This is an important
step for application of the CaT technique
across diverse environments which may have
experienced very different histories of
chemical evolution
\citep[e.g., dwarf galaxies][]{tol03}.

Our results are a striking reaffirmation 
of the utility of the CaT for abundance
measurements of red giants in 
intermediate-age field populations.
Red giants are the 
brightest common stars among old stellar 
populations, and so they are currently 
the only practical targets for abundance
analyses of stars aged $\approx$ 2--14~Gyr
in Local Group galaxies beyond the Milky 
Way halo.  In these systems, the typical
magnitude of the RGB tip is I $\ga$ 20,
and measurement of the CaT from 
low-resolution spectra is the most direct
available method to
accrue the large samples of abundance measurements
needed to assess the chemical evolution
of systems beyond the Milky Way halo.

\section*{Acknowledgments}

We thank K. Venn for suggesting the test described
in \S \ref{mix-sec}, and E. Skillman and S. Trager 
for discussions
and comments on this project.  It is a pleasure
to acknowledge the help of the ESO staff on Paranal,
particularly T. Szeifert, E. Mason, F. Clarke, and
P. Gandhi.
AAC is supported by a Fellowship from the 
Netherlands Research School for Astronomy
(NOVA).
TSH acknowledges financial support from the National
Science Foundation through grant AST-0070985, and
JSG is supported by NSF grant AST-9803018.

{}

\clearpage

\begin{table*}
 \caption{The Cluster Sample}
 \label{clus-tab}
 \begin{tabular}{lccccccl}
  \hline
  Cluster & $\alpha$ (J2000) & $\delta$ (J2000) & [Fe/H] & Age (Gyr) &
          (m$-$M)$_V$ & V$_{\mathrm{HB}}$ & Reference \\
  \hline
  NGC 104 (47 Tuc) & 00:26:33 & $-$71:51 & $-$0.70 $\pm$0.07 & 10.7 & 13.37 & 14.06 & 1,2,3 \\
  NGC 1851         & 05:14:15 & $-$40:04 & $-$0.98 $\pm$0.06 &  9.2 & 15.47 & 16.09 & 2,3,4 \\
  NGC 1904 (M79)   & 05:24:12 & $-$24:31 & $-$1.37 $\pm$0.01 & 11.7 & 15.59 & 16.15 & 1,2,3 \\
  NGC 2298         & 06:48:59 & $-$36:00 & $-$1.74 $\pm$0.06 & 12.6 & 15.59 & 16.11 & 1,2,3 \\
  NGC 4590 (M68)   & 12:39:28 & $-$26:45 & $-$1.99 $\pm$0.10 & 11.2 & 15.19 & 15.68 & 1,2,3 \\
  \hline
  Berkeley 20      & 05:32:34 & $+$00:10 & $-$0.61 $\pm$0.14 &  4.9  & 15.01 & 15.70 & 5,6 \\
  NGC 2141         & 06:03:00 & $+$10:30 & $-$0.33 $\pm$0.10 &  2.8  & 14.16 & 14.90 & 5,7 \\
  Melotte 66       & 07:26:28 & $-$47:41 & $-$0.47 $\pm$0.09 &  6.3  & 13.63 & 14.52 & 5,8 \\
  Berkeley 39      & 07:46:45 & $-$04:41 & $-$0.26 $\pm$0.09 &  7.2  & 13.78 & 14.28 & 5,8 \\
  NGC 2682 (M67)   & 08:51:25 & $+$11:48 & $-$0.15 $\pm$0.05 &  6.3  &  9.98 & 10.54 & 5,8 \\
  \hline
  Trumpler 5       & 06:36:34 & $+$09:27 & $-$0.3:           &  4  : & 14.40 & 15.13 & 9,10 \\
  \hline
 \end{tabular}
 
 \medskip
 References to metallicity, distance modulus, and V$_{\mathrm{HB}}$ values:
 (1) \citet{car97}; (2) \citet{sal02}; (3) \citet{har96}; (4) \citet{rut97b};
 (5) \citet{fri02}; (6) \citet{mac94}; (7) \citet{ros95}; (8) \citet{sar99};
 (9) \citet{kal98}; (10) \citet{kim03}.
\end{table*}

\clearpage

\begin{table}
 \caption{Observed Stars: Globular Clusters}
 \label{glob-tab}
 \begin{tabular}{lccc}
  \hline
  Star  &   V   & V$_{r}$ (km s$^{-1}$) & $\Sigma$W (\AA) \\
  \hline
  \multicolumn{4}{l}{{\bf 47 Tuc}: \citet{lee77}} \\
  L5309 & 12.21 & -21.8 $\pm$7.6 & 7.98 $\pm$0.10  \\
  L5310$^{\ddag}$ & 13.41 & 9.1 $\pm$7.6 & 6.46 $\pm$0.10  \\
  L5312 & 12.18 & -12.3 $\pm$7.5 & 7.91 $\pm$0.10  \\
  L5418 & 15.33 & -20.6 $\pm$7.9 & 5.77 $\pm$0.16  \\
  L5419 & 14.05 & -22.2 $\pm$7.8 & 6.13 $\pm$0.12  \\
  L5422 & 12.47 & -23.1 $\pm$7.6 & 7.38 $\pm$0.10  \\
  L5527 & 13.56 & -25.5 $\pm$7.6 & 6.79 $\pm$0.09  \\
  L5528 & 14.40 & -28.5 $\pm$7.6 & 6.37 $\pm$0.18  \\
  L5530 & 13.14 & -18.6 $\pm$7.6 & 6.92 $\pm$0.08  \\
  \multicolumn{4}{l}{$^{\ddag}$ Radial velocity nonmember, excluded from analysis.}\\
  \hline
  \multicolumn{4}{l}{{\bf NGC 1851}: \citet{ste81}} \\
  003 & 13.60 & 324.5 $\pm$7.4 & 7.52 $\pm$0.15 \\
  065 & 15.75 & 322.1 $\pm$7.5 & 5.67 $\pm$0.08 \\
  095 & 13.57 & 334.1 $\pm$7.4 & 6.94 $\pm$0.11 \\
  107 & 14.50 & 330.7 $\pm$7.5 & 6.78 $\pm$0.13 \\
  109 & 14.85 & 334.3 $\pm$7.4 & 5.96 $\pm$0.10 \\
  112 & 13.80 & 327.3 $\pm$7.4 & 6.85 $\pm$0.11 \\
  123 & 16.21 & 329.7 $\pm$7.4 & 4.99 $\pm$0.08 \\
  126 & 14.29 & 321.5 $\pm$7.5 & 6.95 $\pm$0.11 \\
  160 & 15.51 & 326.4 $\pm$7.7 & 5.85 $\pm$0.09 \\
  175 & 16.22 & 332.3 $\pm$7.9 & 6.07 $\pm$0.10 \\
  179 & 16.45 & 320.4 $\pm$7.8 & 5.66 $\pm$0.08 \\
  195 & 15.82 & 327.6 $\pm$7.5 & 5.75 $\pm$0.09 \\
  209 & 14.08 & 333.4 $\pm$7.6 & 7.05 $\pm$0.20 \\
  231 & 15.93 & 332.8 $\pm$7.8 & 5.89 $\pm$0.10 \\
  275 & 14.98 & 333.8 $\pm$7.6 & 6.79 $\pm$0.13 \\
  \hline
  \multicolumn{4}{l}{{\bf NGC 1904}: \citet{ste77}} \\
  006  & 15.27 & 203.2 $\pm$7.4 & 4.91 $\pm$0.08 \\
  011  & 15.74 & 213.5 $\pm$7.5 & 4.85 $\pm$0.10 \\
  015  & 13.22 & 204.8 $\pm$7.5 & 6.67 $\pm$0.10 \\
  045  & 15.58 & 206.8 $\pm$7.8 & 4.95 $\pm$0.08 \\
  089  & 14.71 & 206.5 $\pm$7.8 & 5.51 $\pm$0.09 \\
  091  & 16.25 & 207.6 $\pm$7.8 & 4.76 $\pm$0.08 \\
  111  & 15.62 & 206.9 $\pm$7.6 & 5.18 $\pm$0.08 \\
  115  & 15.95 & 211.0 $\pm$7.7 & 4.65 $\pm$0.10 \\
  138  & 16.16 & 206.3 $\pm$7.8 & 4.36 $\pm$0.08 \\
  153  & 13.44 & 206.1 $\pm$7.4 & 6.47 $\pm$0.11 \\
  160  & 13.02 & 200.3 $\pm$7.5 & 6.42 $\pm$0.10 \\
  161  & 15.86 & 216.5 $\pm$7.8 & 4.05 $\pm$0.07 \\
  176  & 14.95 & 215.9 $\pm$7.6 & 4.74 $\pm$0.08 \\
  209  & 15.02 & 210.8 $\pm$7.5 & 5.63 $\pm$0.11 \\
  224  & 16.09 & 220.4 $\pm$7.9 & 4.63 $\pm$0.09 \\
  237  & 14.19 & 220.8 $\pm$7.5 & 5.77 $\pm$0.11 \\
  241  & 13.61 & 228.5 $\pm$7.5 & 5.80 $\pm$0.09 \\
  \hline
  \end{tabular}
  \end{table}
  \begin{table}
  \contcaption{Observed Stars: Globular Clusters}
  \begin{tabular}{lccc}
  \hline
   Star  &   V   & V$_{r}$ (km s$^{-1}$) & $\Sigma$W (\AA) \\
  \hline
  \multicolumn{4}{l}{{\bf NGC 2298}: \citet{alc86}} \\
  6     & 13.82 & 156.5 $\pm$7.6 & 4.97 $\pm$0.07  \\
  12    & 14.23 & 146.6 $\pm$7.5 & 4.84 $\pm$0.07  \\
  15    & 14.86 & 151.2 $\pm$7.6 & 4.31 $\pm$0.06  \\
  22    & 15.30 & 162.0 $\pm$7.4 & 4.06 $\pm$0.06  \\
  25    & 15.69 & 153.2 $\pm$7.5 & 4.15 $\pm$0.06  \\
  S156$^{\star}$  & 15.53 & 148.5 $\pm$7.6 & 4.14 $\pm$0.05  \\
  S172$^{\star}$  & 15.61 & 160.6 $\pm$7.7 & 3.97 $\pm$0.05  \\
  \multicolumn{4}{l}{$^{\star}$ ID and magnitude from T. Smecker-Hane, 
   unpublished.} \\
  \hline
  \multicolumn{4}{l}{{\bf NGC 4590}: \citet{har75}} \\
  I-2   & 14.95 &  -96.0 $\pm$7.6 & 2.98 $\pm$0.04   \\
  I-49  & 14.62 &  -93.6 $\pm$7.8 & 3.04 $\pm$0.05   \\
  I-74  & 14.59 & -103.7 $\pm$7.7 & 3.21 $\pm$0.05   \\
  I-119 & 13.62 &  -89.1 $\pm$7.6 & 3.78 $\pm$0.06   \\
  I-239 & 14.19 &  -86.0 $\pm$7.6 & 3.63 $\pm$0.05   \\
  I-256 & 12.64 &  -90.9 $\pm$7.6 & 5.20 $\pm$0.08   \\
  I-258 & 14.24 &  -90.2 $\pm$7.7 & 3.14 $\pm$0.05   \\
  I-260 & 12.52 &  -91.4 $\pm$7.5 & 4.85 $\pm$0.07   \\
  II-47 & 15.03 &  -85.8 $\pm$7.8 & 3.19 $\pm$0.05   \\
  \hline
 \end{tabular}
\end{table}

\clearpage

\begin{table}
 \caption{Observed Stars: Open Clusters}
 \label{open-tab}
 \begin{tabular}{lccc}
  \hline
  Star  &   V   & V$_{r}$ (km s$^{-1}$) & $\Sigma$W (\AA ) \\
  \hline
  \multicolumn{4}{l}{{\bf Berkeley 20}: \citet{mac94}} \\
  05 & 14.80 & 73.9 $\pm$7.5 & 7.56 $\pm$0.12  \\
  08 & 15.15 & 79.2 $\pm$7.5 & 7.41 $\pm$0.12  \\
  12 & 16.21 & 83.8 $\pm$7.5 & 6.61 $\pm$0.12  \\
  22 & 16.90 & 74.6 $\pm$7.5 & 6.19 $\pm$0.13  \\
  29$^{\ddag}$ & 17.14 & 42.1 $\pm$7.7 & 5.52 $\pm$0.17  \\
  \multicolumn{4}{l}{$^{\ddag}$ Radial velocity nonmember, excluded from analysis.}\\
  \hline
  \multicolumn{4}{l}{{\bf NGC 2141}: \citet{bur72,ros95}} \\
  1-3-21  & 14.18 &  31.8 $\pm$7.5 & 8.04 $\pm$0.18  \\
  1-4-05  & 14.61 &  50.7 $\pm$7.5 & 7.58 $\pm$0.13  \\
  3-2-18  & 13.05 &  26.0 $\pm$7.5 & 9.35 $\pm$0.18  \\
  3-2-34  & 14.95 &  27.6 $\pm$7.5 & 7.26 $\pm$0.17  \\
  3-2-40  & 13.25 &  28.9 $\pm$7.4 & 8.80 $\pm$0.17  \\
  3-2-52  & 14.36 &  33.2 $\pm$7.5 & 7.67 $\pm$0.14  \\
  3-2-56$^{\ddag}$  & 15.54 & -11.3 $\pm$7.5 & 4.15 $\pm$0.09  \\
  4-08    & 14.80 &  23.2 $\pm$7.6 & 7.76 $\pm$0.13  \\
  4-09    & 13.27 &  28.7 $\pm$7.6 & 8.78 $\pm$0.14  \\
  4-13    & 15.15 &  32.1 $\pm$7.4 & 6.79 $\pm$0.14  \\
  4-14    & 15.52 &  28.8 $\pm$7.4 & 7.38 $\pm$0.12  \\
  4-24    & 14.77 &  31.4 $\pm$7.5 & 7.64 $\pm$0.13  \\
  4-25    & 14.13 &  31.3 $\pm$7.5 & 8.11 $\pm$0.18  \\
  5-09    & 14.62 &  33.0 $\pm$7.5 & 7.69 $\pm$0.14  \\
  5-13    & 13.96 &  32.0 $\pm$7.5 & 7.92 $\pm$0.15  \\
  5-13a   & 15.00 &  49.3 $\pm$7.4 & 7.00 $\pm$0.16  \\
  \multicolumn{4}{l}{$^{\ddag}$ Radial velocity nonmember, excluded from analysis.}\\
  \hline
  \multicolumn{4}{l}{{\bf Melotte 66}: \citet{ant79}} \\
  1205 & 14.18 & 18.3 $\pm$7.5 & 6.86 $\pm$0.11   \\
  2107 & 14.70 & 14.9 $\pm$7.5 & 6.57 $\pm$0.11   \\
  2133 & 13.20 & 10.9 $\pm$7.5 & 7.91 $\pm$0.12   \\
  2226 & 14.08 & 18.9 $\pm$7.5 & 7.11 $\pm$0.19   \\
  2233 & 15.44 & 14.4 $\pm$7.4 & 6.51 $\pm$0.09   \\
  2244 & 14.47 & 16.4 $\pm$7.6 & 6.97 $\pm$0.09   \\
  2261 & 13.64 & 41.2 $\pm$7.5 & 7.46 $\pm$0.16   \\
  3101 & 14.98 & 15.6 $\pm$7.7 & 5.03 $\pm$0.10   \\
  3133 & 14.08 & 17.2 $\pm$7.6 & 6.98 $\pm$0.12   \\
  3229 & 13.93 &  7.7 $\pm$7.5 & 7.08 $\pm$0.31   \\
  3235 & 14.66 & 18.3 $\pm$7.7 & 6.89 $\pm$0.10   \\
  3260 & 14.56 & 16.2 $\pm$7.4 & 6.90 $\pm$0.12   \\
  4151 & 12.69 & 12.7 $\pm$7.6 & 8.23 $\pm$0.13   \\
  4265 & 14.07 &  7.4 $\pm$7.5 & 6.97 $\pm$0.12   \\
  \hline
  \end{tabular}
  \end{table}
  \begin{table}
  \contcaption{Observed Stars: Open Clusters}
  \begin{tabular}{lccc}
  \hline
  Star  &   V   & V$_{r}$ (km s$^{-1}$) & $\Sigma$W (\AA) \\
  \hline\multicolumn{4}{l}{{\bf Berkeley 39}: \citet{kal89}} \\
  002 & 13.09 & 56.5 $\pm$7.5 & 8.39 $\pm$0.18 \\
  003 & 13.55 & 61.0 $\pm$7.4 & 8.26 $\pm$0.18 \\
  005 & 13.88 & 56.5 $\pm$7.5 & 7.68 $\pm$0.16 \\
  009 & 14.23 & 56.7 $\pm$7.5 & 7.14 $\pm$0.13 \\
  012 & 14.34 & 55.9 $\pm$7.4 & 7.37 $\pm$0.14 \\
  013 & 14.35 & 58.2 $\pm$7.5 & 6.87 $\pm$0.16 \\
  016 & 14.48 & 61.4 $\pm$7.5 & 7.28 $\pm$0.15 \\
  017 & 14.75 & 54.0 $\pm$7.6 & 6.72 $\pm$0.12 \\
  018 & 14.76 & 57.9 $\pm$7.5 & 6.88 $\pm$0.14 \\
  028 & 15.40 & 58.9 $\pm$7.5 & 6.60 $\pm$0.15 \\
  \hline
  \multicolumn{4}{l}{{\bf M67}$^{\star}$: \citet{joh55,san77}} \\
  F104       & 11.20 &  33.5 & 7.13 $\pm$0.16  \\
  F105       & 10.30 &  34.3 & 8.00 $\pm$0.16  \\
  F108       &  9.72 &  34.7 & 8.36 $\pm$0.17  \\
  F135       & 11.44 &  34.3 & 7.10 $\pm$0.13  \\
  F141       & 10.48 &  33.6 & 7.73 $\pm$0.14  \\
  F164       & 10.55 &  33.3 & 7.40 $\pm$0.15  \\
  F170       &  9.69 &  34.3 & 8.28 $\pm$0.19  \\
  1264$^{\S}$ & 11.74 & 34.9 & 5.83 $\pm$0.23  \\
  \multicolumn{4}{l}{$^{\star}$ Radial velocities from \citet{mat86}.}\\
  \multicolumn{4}{l}{$^{\S}$ Blend \citep*{mon93}; excluded from analysis.} \\
  \hline
 \end{tabular}
\end{table}

\clearpage

\begin{table}
 \caption{Observed Stars: Trumpler 5}
 \label{tr5-tab}
 \begin{tabular}{lcccc}
  \hline
  Star$^{\dagger}$ & V (mag) & V$_{r}$ (km s$^{-1}$) &
  $\Sigma$W (\AA ) & Note \\
  \hline
  K3763 & 14.54 & -3.9 $\pm$7.5 & 7.71 $\pm$0.03 & 1 \\
  K3354 & 14.48 & 53.7 $\pm$7.5 & 6.85 $\pm$0.02 &   \\
  K3066 & 14.39 & 50.7 $\pm$7.4 & 7.04 $\pm$0.03 &   \\
  K2579 & 15.59 & 47.8 $\pm$7.5 & 6.37 $\pm$0.03 &   \\
  K2565 & 15.79 & 60.5 $\pm$7.5 & 6.12 $\pm$0.03 &   \\
  K2324 & 15.03 & 53.1 $\pm$7.5 & 6.70 $\pm$0.02 &   \\
  K2280 & 15.16 & 47.8 $\pm$7.4 & 6.54 $\pm$0.03 &   \\
  K1935 & 12.87 & 53.5 $\pm$7.5 & 8.63 $\pm$0.03 & 2 \\
  K1834 & 16.24 & 50.8 $\pm$7.7 & 6.21 $\pm$0.05 &   \\
  K1401 & 15.29 & 14.8 $\pm$7.5 & 6.60 $\pm$0.03 & 1 \\
  K1305 & 14.24 & 64.9 $\pm$7.5 & 8.12 $\pm$0.03 & 3 \\
  K1277 & 14.92 & 46.3 $\pm$7.2 &                & 4 \\
  K1214 & 13.91 & 55.7 $\pm$7.5 & 7.61 $\pm$0.03 & 2 \\
  K1026 & 14.73 & 52.0 $\pm$7.3 & 6.48 $\pm$0.10 &   \\
   K833 & 14.91 & 54.4 $\pm$7.4 & 6.53 $\pm$0.04 & 3 \\
   K488 & 16.51 & 57.1 $\pm$7.7 & 7.31 $\pm$0.05 & 3 \\
  \hline
 \end{tabular}

 \medskip
 $^{\dagger}$ From Table 3 of Ka\l uzny (1998).\\
 Notes: (1) radial velocity nonmember; (2) weak TiO?;
 (3) Probably differentially reddened; (4) Carbon 
 star $\equiv$ V493~Mon.
\end{table}

\begin{table}
 \caption{Cluster Radial Velocities}
 \label{vrad-tab}
 \begin{tabular}{lcccc}
 \hline
          & V$_r$ & $\sigma$(V$_r$) & prev.\ V$_r$ &   \\
 Cluster  & (km s$^{-1}$) & (km s$^{-1}$) & (km s$^{-1}$) & Reference \\   
 \hline
 47 Tuc   & $-$22 & 5 & $-$18.7 & 1 \\
 NGC 1851 &   329 & 5 &   320.5 & 1 \\
 NGC 1904 &   211 & 7 &   206.0 & 1 \\
 NGC 2298 &   154 & 6 &   148.9 & 1 \\
 NGC 4590 & $-$92 & 6 & $-$94.3 & 1 \\
 \hline
 Be 20    &    78 & 5 &      70 & 2 \\
 NGC 2141 &    33 & 5 &      64 & 3 \\
 Mel 66   &    16 & 8 &      23 & 4 \\
 Be 39    &    58 & 2 &      55 & 2 \\
 M67      &       &   &      33 & 2 \\
 \hline
 Tr 5     &    54 & 5 &         &   \\
 \hline
 \end{tabular}

 \medskip
 References: (1) \citet{har96}; (2) \citet{fri02};
 (3) \citet*{fri89}; (4) \citet{fri93}.
\end{table}

\clearpage

\begin{table*}
 \caption{Cluster Metallicity Data}
 \label{met-tab}
 \begin{tabular}{lcccccc}
 \hline
          & W$^{\prime}$   &    [Fe/H]         &       [Fe/H]          & [Fe/H]$^{\dag}$        &       [Fe/H]         &           \\
 Cluster  &    (\AA )      &    (ref)          & (Eqn. \ref{cal-eqn})  &     (alt.)             & (Eqn. \ref{alt-cal}) & Reference \\
 \hline
 47 Tuc   & 6.36 $\pm$0.03 & $-$0.70 $\pm$0.07 & $-$0.66 $\pm$0.09     & $-$0.88, {\it $-$0.71} & $-$0.77 $\pm$0.13    & 1,{\it 2} \\
 NGC 1851 & 5.55 $\pm$0.02 & $-$0.98 $\pm$0.06 & $-$0.96 $\pm$0.12     & $-$1.19, {\it $-$1.33} & $-$1.17 $\pm$0.16    & 1,{\it 2} \\
 NGC 1904 & 4.40 $\pm$0.02 & $-$1.37 $\pm$0.01 & $-$1.37 $\pm$0.11     & $-$1.64, {\it $-$1.68} & $-$1.66 $\pm$0.12    & 1,{\it 2} \\
 NGC 2298 & 3.54 $\pm$0.03 & $-$1.74 $\pm$0.06 & $-$1.69 $\pm$0.07     & $-$2.07, {\it $-$1.81} & $-$1.96 $\pm$0.06    & 1,{\it 2} \\
 NGC 4590 & 2.48 $\pm$0.03 & $-$1.99 $\pm$0.10 & $-$2.07 $\pm$0.09     & $-$2.43, {\it $-$2.09} & $-$2.25 $\pm$0.26    & 1,{\it 2} \\
 Be 20    & 6.91 $\pm$0.03 & $-$0.61 $\pm$0.14 & $-$0.47 $\pm$0.07     & {\it $-$0.4}           & $-$0.47 $\pm$0.11    & {\it 3}   \\
 NGC 2141 & 7.47 $\pm$0.02 & $-$0.33 $\pm$0.10 & $-$0.26 $\pm$0.10     & {\it $-$0.06}          & $-$0.14 $\pm$0.16    & {\it 4}   \\
 Mel 66   & 6.86 $\pm$0.03 & $-$0.47 $\pm$0.09 & $-$0.48 $\pm$0.06     & $-$0.38, {\it $-$0.35} & $-$0.50 $\pm$0.10    & 5,{\it 4} \\
 Be 39    & 7.32 $\pm$0.03 & $-$0.26 $\pm$0.09 & $-$0.32 $\pm$0.09     & {\it $-$0.18}          & $-$0.24 $\pm$0.14    & {\it 4}   \\
 M67      & 7.68 $\pm$0.06 & $-$0.15 $\pm$0.05 & $-$0.19 $\pm$0.05     & $-$0.05, {\it 0.00}    & $-$0.01 $\pm$0.09    & 6,{\it 4} \\
 Tr 5     & 6.63 $\pm$0.03 &                   & $-$0.56 $\pm$0.11     & {\it $-$0.3}           & $-$0.62 $\pm$0.14    & {\it 7}   \\
 \hline
 \end{tabular}

 \medskip
 $^{\dag}$ Values in roman type are from high-dispersion spectroscopy; values
     in {\it italic} type are from photometric or spectrophotometric indices.\\
 References: (1) \citet{kra03}; (2) \citet{zin84}; (3) \citet{dur01};
   (4) \citet{twa97}; (5) \citet{gra94}; (6) \citet{she00}; (7) \citet{kim03}.
\end{table*}

\clearpage

\clearpage 

\clearpage

\clearpage

\clearpage

\end{document}